# NFV: State of the Art, Challenges and Implementation in Next Generation Mobile Networks (vEPC)

Hassan Hawilo[1], Abdallah Shami[1], Maysam Mirahmadi[1,2,] and Rasool Asal[3]

[1]Department of Electrical and Computer Engineering, Western University, Canada

[2]IBM Canada Ltd

[3] British Telecom, UK

*Abstract*— **As mobile network users look forward to the connectivity speeds of 5G networks, service providers are facing challenges in complying with connectivity demands without substantial financial investments. Network Function Virtualization (NFV) is introduced as a new methodology that offers a way out of this bottleneck. NFV is poised to change the core structure of telecommunications infrastructure to be more cost-efficient. In this paper, we introduce a Network Function Virtualization framework, and discuss the challenges and requirements of its use in mobile networks. In particular, an NFV framework in the virtual environment is proposed. Moreover, in order to reduce signaling traffic and achieve better performance, this paper proposes a criterion to bundle multiple functions of virtualized evolved packet-core in a single physical device or a group of adjacent devices. The analysis shows that the proposed grouping can reduce the network control traffic by 70 percent.**

## I. Introduction

The demand for reducing capital expenditures (CAPEX) and operating expenditures (OPEX) has pushed information technology (IT) specialists toward contemplating designs to achieve more effective capital investments with higher return on capital. Toward this goal, the virtualization technology has emerged as a way to decouple software applications from the underlying hardware and enable software to run in a virtualized environment. In a virtual environment, hardware is emulated, and the operating system (OS) runs over the emulated hardware as if it is running on its own bare-metal resources. Using this procedure, multiple virtual machines can share available resources and run simultaneously on a single physical machine [1].

The demand for broadband network connectivity has been increasing dramatically in the last decade. It gains additional momentum with the increase in the number of Internet-connected mobile devices, ranging from smartphones, tablets, and laptops to sensor networks, and machine-to-machine (M2M) services. This increasing demand is pushing network service providers to invest in infrastructure to keep up with the demand, although studies show that the return on such investments is minimal [2]. Network expenditures depend highly on the infrastructure on which the network relies. The high cost of any network-improvement upgrade or new service release narrows the revenue margin of the service provider. Network operating challenges are not limited to the cost of expensive hardware devices, but also include increasing energy costs and the competitive market for highly qualified personnel with the skills necessary to design, integrate, and operate an increasingly complex hardware-based infrastructure. In addition, managing network infrastructure is another major concern of service providers. These issues do not affect revenue only, but they also increase time-to-market and limit innovation in the telecommunications industry. Therefore, network operators seek to minimize or even eliminate their dependency on proprietary hardware.

To achieve these targets successfully, a group of seven telecom operators have formed an industry specifications group for Network Function Virtualization (NFV) under the European Telecommunications Standards Institute (ETSI). They revealed their solution in October 2012 [3]. More recently, several telecom-equipment providers and IT specialists joined the group.

This paper introduces NFV and provides a guideline to design and implement it in the core network of telecommunication networks. Section II gives a conceptual definition of NFV. Section III discusses the NFV framework. In Section IV, an approach for implementing NFV framework entities is provided. In Section V, the challenges and requirements of NFV are discussed. Section VI defines different services and use-cases of NFV in mobile networks. The virtualized evolved packet core (vEPC) network is discussed in Section VII. Section VIII introduces the proposed grouping approach for EPC entities in the NFV environment. Finally, Section XI concludes the paper.

## II. Network Function Virtualization

The substantial dependence of networks on their underlying hardware and the existence of various specialized hardware appliances, for example firewalls, deep packet inspection (DPI) equipment, and routers, in the network infrastructure have escalated the challenges facing network service providers. Furthermore, the reduced life cycles of these types of hardware due to fast pace of innovation tends to multiply CAPEX and OPEX investments [3]. Network function virtualization technology was developed to take advantage of the evolution of IT virtualization. It separates network functions from the underlying proprietary hardware appliances. NFV is the concept of transferring network functions from dedicated hardware appliances to software-based applications running on commercial off-the-shelf (COTS) equipment. These applications are executed and consolidated on standard IT platforms like high-volume servers, switches, and storage. Through NFV, network functions can be instantiated in various locations such as datacenters, network nodes, and end-user premises as the network requires [3].

NFV provides many benefits to the telecommunications industry. Some of these benefits are openness of platforms, scalability and flexibility, operating performance improvement, shorter development cycles, and reduced CAPEX and OPEX investments [3].

### III. NFV FRAMEWORK

The basic components of virtualized platforms where NFV is deployed are: a) Physical server: The physical server is the bare-metal machine that has all the physical resources such as CPU, storage, and RAM. b) Hypervisor: The hypervisor, or virtual machine monitor, is the software that runs and manages physical resources. It provides the virtual environment on which the guest virtual machines are executed. c) The guest virtual machine: A piece of software that emulates the architecture and functionalities of a physical platform on which the desired application is executed.

Virtual machines (VMs) are deployed on high-volume servers which can be located in datacenters, at network nodes, and in end-user facilities. Moreover, most VMs provide on-demand computing resources using cloud. Cloud-computing services are offered in various formats [4]: infrastructure as a service (IaaS) that is also referred to as hardware as a service (HaaS), platform as a service (PaaS), software as a service (SaaS), and network as a service (NaaS). There is no agreement on a standard definition of NaaS. However it is often considered to be provided under IaaS. The NFV technology takes advantage of infrastructure and networking services (IaaS and NaaS) to form the network function virtualization infrastructure (NFVI) [5].

To achieve the objectives promised by NFV, such as flexibility in assigning virtual network functions (VNFs) to hardware, rapid service innovation, enhanced operational efficiency, reduced power usage, and open standard interfaces between VNFs, each VNF should run on a framework that includes dynamic initiation and orchestration of VNF instances. In addition, it should also manage the NFVI hosting environment on IT virtualization technologies to meet all VNF requirements regarding data, resource allocation, dependencies, availability, and other attributes. The ETSI NFV group has defined the NFV architectural framework at the functional level using functional entities and reference points, without any indication of a specific implementation. The functional entities of the architectural framework and the reference points are listed and defined in [5] and shown in Figure 2.

### IV. PROPOSED PLACEMENT OF FRAMEWORK ENTITIES

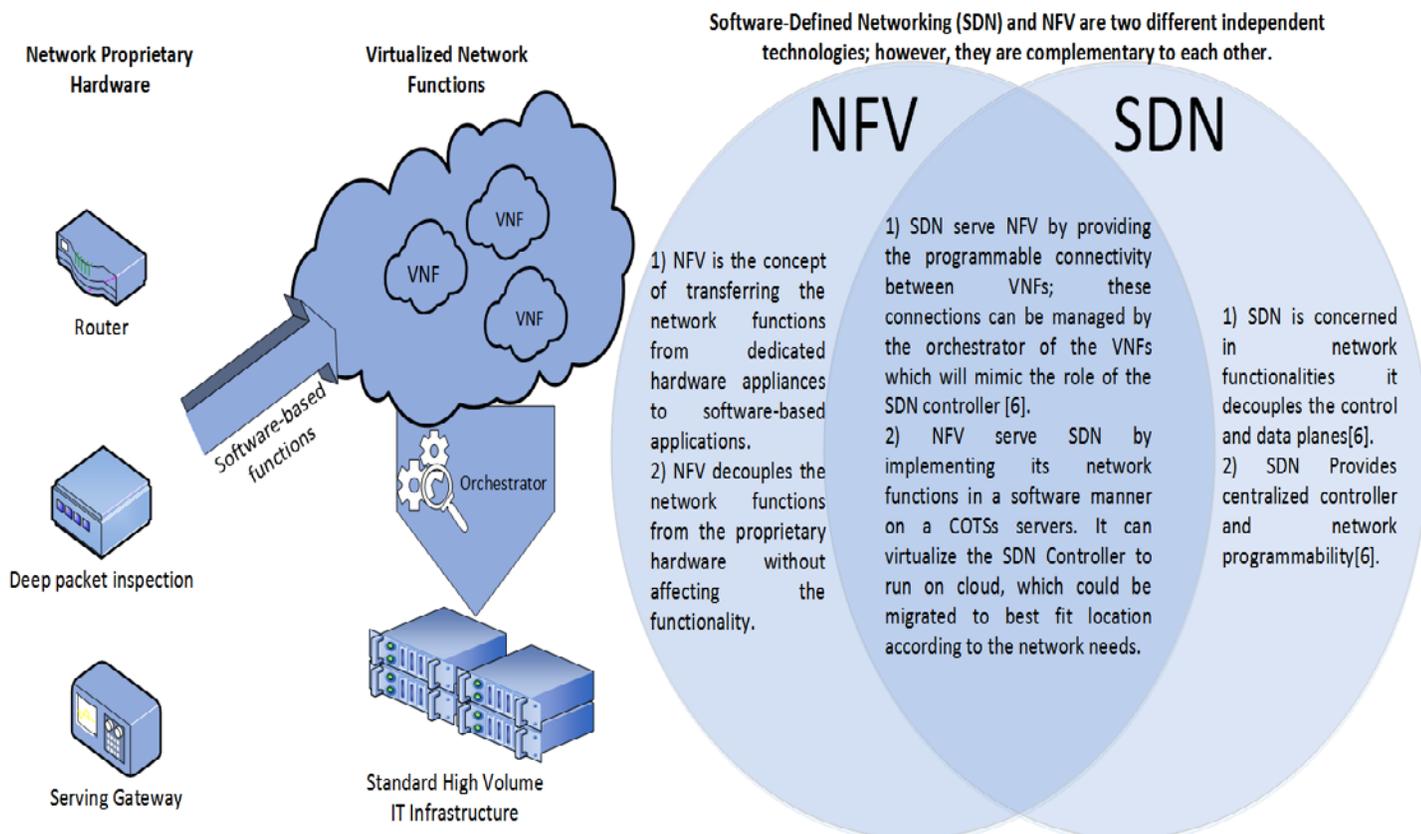

*Figure 1: Network Function Virtualization Concept; NFV differs from SDN*

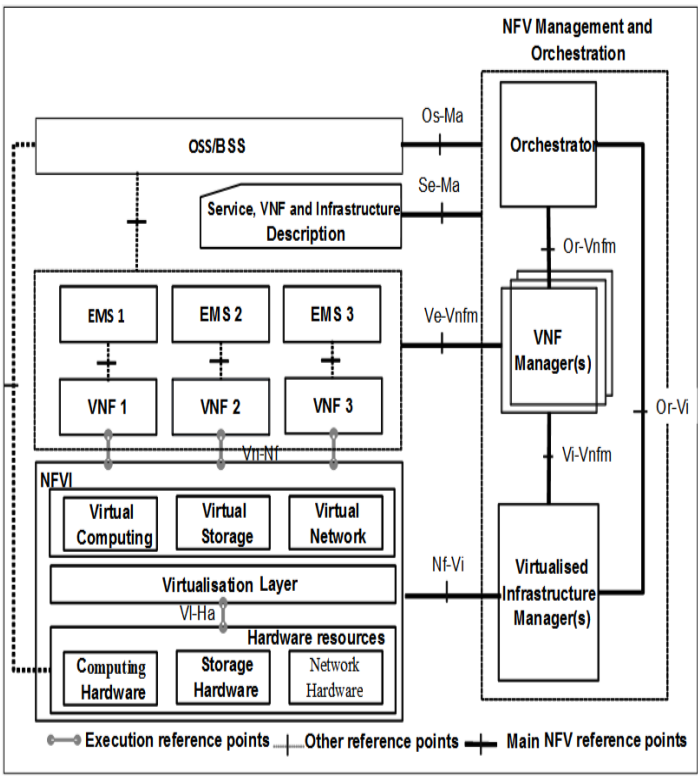

*Figure 2: NFV Framework [5].*

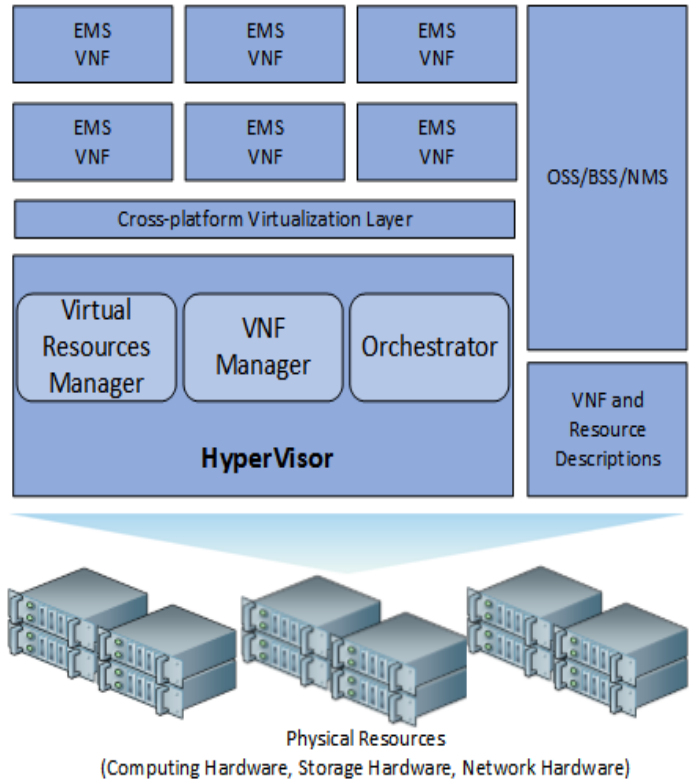

Figure 3: NFV Framework Entities Proposed Placements.

The proposed placement is based on mapping the NFV framework entities to best fit into the virtual environment. The Virtual Resources Manager, the VNF Manager, and the Orchestrator have been grouped at the hypervisor level. Since the virtual environment will not only host VNFs, but also other IT applications, this grouping leads to a centralized controller. The infrastructure that provides NFVI as a service provide cloud services simultaneously on the same hardware resources. Essentially, the hypervisor manages and orchestrates the physical and logical resources of the virtualized environment. It is aware of the virtual machines that are using the underlying hardware and manages resource scheduling and decisions such as migration, resource scaling, and fault and failure recovery, more efficiently to meet the specified quality-of-service requirements of VMs (VNFs and APPs) [5].

The virtualization layer consists of a cross-platform virtual resource manager that runs on top of the hypervisor to ensure the portability and flexibility of VNF independently of the hypervisor. OpenStack, Eucalyptus, oVirt, OpenNebula, and Nimbula are examples of cross-platform virtual layers [7].

The virtual machine hosts VNF and its element-management system (EMS). Each VNF instance has its private EMS to reduce complexity when migrating an existing VNF or initiating a new one.

Operations and business support systems with VNF infrastructure description entities are deployed in a centralized form which provides uniformity of VNF software images and minimize database fragmentation. The proposed placements are illustrated in Figure 3.

## V. RECEPTION BY INDUSTRY

Service providers have shown keen interest in NFV. Observing this interest, telecom equipment vendors and IT companies have started to investigate different aspects of NFV realization. Leading telecommunication equipment vendors like Ericsson, Nokia, Alcatel-Lucent, and Huawei have already started to adopt and upgrade their equipment to support NFV [8-9]. Moreover, leading IT companies that provide carrier grade software like Wind River, 6wind, Qosmos, and HP have been working closely with Intel to optimize their software on Intel processors in order to achieve higher packet processing computations that enable NFV and SDN on COTS platforms. Intel has released the Data Plane Development Kit (DPDK) and has scheduled the release of signal processing development kit in its software development roadmap to extend and speed up NFV and SDN adoption [10]. Service providers started experimenting with these NFV products and put the devices under heavy testing to ensure that they will meet the expectations as carrier-grade products [11].

## VI. NFV CHALLENGES AND REQUIREMENTS

Although NFV is a promising solution for telecommunications service providers, it faces certain challenges that could degrade its performance and hinder its implementation in the telecommunications industry. In this section, some of the NFV requirements and challenges, and proposed solutions are discussed. Table 1 summarizes this section.

### A. Security

Security is an important aspect of the telecommunications industry. NFV should obtain a security level close to that of a proprietary hosting environment for network functions. The best way to achieve this security level is by dividing it according to functional domains. Security in general can be defined according to the following functional domains:
1) Virtualization environment domain (hypervisor)
2) Computing domain
3) Infrastructure domain (networking)
4) Application domain.

Security attacks are expected to increase when implementing network functions in a virtualization environment. A protected hypervisor should be used to prevent any unauthorized access or data leakage. Moreover, other processes such as data communication and VM migration should run in a secure environment [12]. NFV uses

APIs to provide programmable orchestration and interaction with its infrastructure. These APIs introduce a higher security threat on VNFs [13]. The security challenges and the proposed solutions are listed in Table 1.

### B. Computing Performance

The virtual environment underlying hardware server characteristics such as processor architecture, clock rate, cache memory size, memory bandwidth, and speed has a profound impact on VNF performance. VNF software design also plays a major role in VNF performance. VNF software can achieve high performance using the following techniques:
1) A high-demand VNF should be implemented using multi-threading techniques and in a distributed and scalable fashion, in order to execute it on multiple cores or different hosts.
2) Software instances should have independent memory structures to avoid operating-system deadlocks.
3) VNF should implement its own network stack and avoid networking stacks implementation in the operating system, which consume large amounts of computing resources.
4) Direct access to input/output interfaces should be used

| Challenge | Description | Solutions and Requirements |
|---|---|---|
| Security | Virtualization security risks according to functional domains:<br><br>1) Virtualization environment domain (Hypervisor):<br>  – Unauthorized access or data leakage.<br>2) Computing domain:<br>  – Shared computing resources: CPU, memory…etc.<br>3) Infrastructure domain (networking):<br>  – Shared logical-networking layer (Vswitches).<br>  – Shared physical NICs. | Security implementations according to functional domains:<br><br>1) Virtualization environment domain (Hypervisor):<br>  – Isolation of the served virtual-machine space, with access provided only with authentication controls.<br>2) Computing domain:<br>  – Secured threads.<br>  – Private and shared memory allocations should be erased before their re-allocation.<br>  – Data should be used and stored in an encrypted manner by which exclusive access is provided only to the VNF.<br>3) Infrastructure domain (networking):<br>  – Usage of secured networking techniques (TLS, IPSec, or SSH). |
| Computing performance | The virtualized network function should provide comparable performance to network functions running on proprietary hardware equipment. | VNF software could achieve high performance using the following techniques:<br>  – Multithreading to be executed over multiple cores, or could be scaled over different hosts.<br>  – Independent memory structures to avoid operating-system deadlocks.<br>  – VNF should implement its own network stack.<br>  – Direct access to input/output interfaces.<br>  – Processor affinity techniques should be implemented. |
| VNF interconnection | Virtualized environment has different approaches from classical network function interconnection. | VNFs should take advantage of accelerated Vswitches and use NICs that are single-root I/O virtualization (SR-IOV) compliant. |
| Portability | VNFs should be decoupled from any underlying hardware and software. VNFs should be deployable on different virtual environments to take advantage of virtualization techniques like live migrations. | The VNF development should be based on a cross-platform virtual resource manager that ensure its portability. |
| Operation and management<br>Existence with legacy networks<br>Carrier-grade service assurance | VNFs should be easy to manage and migrate with existing legacy systems without losing the specification of a carrier-grade service. | To achieve the desired operation and management performance, a standard template of NFV framework entities should be well-defined. It should be able to interact with legacy management systems with minimal effects on existing networks. The NFV orchestrator must monitor network function performance almost in real time. |

*Table 1: NFV Challenges and Solutions.*

whenever possible to reduce latency and increase data throughput.

5) Processor affinity techniques should be used to take advantage of cache memories.

Implementing these techniques in VNF software may require a different approach from the automated resource allocation within a given pool of servers, which is currently used in IT environments.

### C. Interconnection of VNFs

Unlike the classical approach of interconnecting network functions by a direct connection or through Layer 2 (L2) switches, a virtualized environment uses different approaches. In a virtualized environment, virtual machines can be connected in different scenarios [14]:

1) If two VNFs are on the same physical server and on the same local-access network (LAN), they would be connected through the same Vswitch.

2) If two VNFs are on the same physical server but on different LANs, the connection passes through the first Vswitch to the network interface controller (NIC), then to the external switch, and back again to the same NIC. This NIC forwards the connection to the Vswitch of the second LAN and then to the VNF.

3) If two VNFs are on different servers, the connection passes through the first Vswitch to the NIC and then to an external switch. This switch forwards the connection to the NIC of the desired server. Finally, this NIC forwards it to its internal Vswitch and then to the destination VNF.

Some NICs provide direct access from the virtual machine. These NICs are single-root I/O virtualization (SR-IOV) compliant. They offer faster and higher throughput to virtual machines. Each connectivity technique has its own advantages in terms of performance, flexibility, and isolation. Virtual interfaces managed by the hypervisor have lower performance compared to virtual interfaces offered by SR-IOV-compliant NICs. However, virtual interfaces provided by the hypervisor are simpler to configure, and support VM live migration in a simpler way. The right choice depends on VNF workloads.

### D. Portability

Virtualized network functions can be deployed in different ways. Each way has its own advantages and drawbacks. Virtualized network functions that are executed directly on bare-metal ensure predictable performance because mappings of software instances to hardware are predictable. This kind of deployment sacrifices resource isolation and makes software-instance security difficult to achieve because multiple software appliances are executed as processes on the same operating system. In addition, the designed software would be OS-dependent.

Deploying virtual network functions through a virtual environment improves portability and ensures that hardware resources are viewed uniformly by the VNF. This deployment also enables each VNF to be executed on its specific operating system while remaining unaware of the underlying operating system. In addition, VNF resource isolation is ensured because

VNFs are executed on independent VMs managed by the hypervisor layer, which guarantees no unexpected interactions between them. Strict mapping of resources should be used to guarantee resource isolation.

### E. Operation and Management

Virtual network functions should be implemented as simple drag-and-drop operations in the orchestration management system. To make this a reality, both VNFs and computing infrastructure should be described using standard templates that enable automated management.

The orchestration management system is responsible for providing and managing the NFV environment through north- and south-bound interactions. North-bound interactions are used to manage and provide access to the VNFs. Moreover, VNFs could use them for information or request queries such as asking for more computing resources. South-bound interactions are used to interact with the NFVI and request information from other framework entities. In addition, they are used to request information about policies, VNF software images, VNF descriptions, or network forwarding graphs.

### F. Co-existence with Legacy Networks

Virtual network functions should be able to coexist with legacy network equipment. It means that a) it should be able to interact with legacy management systems with minimal effects on existing networks, b) the network forwarding graph should not be affected by the existence of one or more VNFs, and c) a secured transition should be ensured between VNF instances and physical functions, without any service interruption or performance impacts [15].

### G. Carrier-Grade Service Assurance

Carrier-grade service is a service in which hardware, software, and system components ensure high availability and reliability. For NFV to meet carrier-grade service requirements, it should provide resilience to failure, service continuity, and service assurance. Resilience to failure is provided by implementing an automated on-demand mechanism in the NFV framework to reconstitute the VNF after a failure. VNF reconstitution should not have any impact on the system to ensure stable service. Service assurance is provided by the NFV orchestrator, which is monitoring network-function performance and scale resources almost in real time [16].

## VII. USE CASES AND SERVICES

The Network Function Virtualization technology in principle considers all network functions for virtualization through well-defined standards. Most likely, NFV services will be provided in a similar way to IT virtualization service models. NFV service models include NFVI as a Service (NFVIaaS), VNF as a Service (VNFaaS), and Virtual Network Platform as a Service (VNPaaS). Service providers will choose between these service models to serve their network-connectivity needs and use cases. Some of the use cases will include, for example, fixed-access network function virtualization, content-delivery network virtualization, and home environment virtualization [17].

### A. Mobile network virtualization

Mobile network connectivity demand is rapidly increasing with the growing number of mobile devices and applications that need to be always connected. Service operators must continually upgrade and enhance their infrastructure, for example by providing enough mobile base stations and network cores to achieve the desired data throughput, latency, and quality of service. Virtualization of mobile networks targets the mobile-network base station and mobile core network. Service providers have been showing interest in virtualizing mobile base stations so that they can consolidate as many network functions as possible in a standard hardware as needed to serve different mobile network technologies with a single virtualized mobile base station. Virtualizing the mobile base station is challenging because it hosts signal-processing functions in its physical layer. Therefore, virtualization first is considered for implementation in the higher network stack layers. Considering eNodeB, which is the fourth-generation network (LTE) base station, virtualization will be implemented in layer 3 and then in layer 2 [17]. Layer 3 hosts the functionalities of the control and data plane that connects to the mobile core network. Layer 2 hosts the packet data convergence protocol (PDCP), radio link control (RLC), and media access control (MAC) network functions. Virtualizing layer 2 and 3 of the base station provide the opportunity to offer a centralized computing infrastructure for multiple base stations, which lead to lower-cost base stations because only baseband signal processing should be implemented on-site. Furthermore, service providers will benefit from sharing their remote base-station infrastructure to achieve better area coverage with minimum CAPEX and OPEX investment. There is also some efforts to centralize the L1 functionalities of several base stations [18]. They will be able to upgrade VNFs to support multiple telecommunications technologies and adapt them for new releases.

## VIII. VIRTUALIZATION OF THE EVOLVED PACKET CORE (EPC)

The mobile core network is the most important part of the network in many access technologies. Virtualizing the functionalities within the core is the main target for NFV. The most recent core network is the evolved packet core (EPC) network. EPC has been introduced in release 8 as a simplified all-IP core network architecture. It is designed to permit mobile broadband services by combining leading-edge IP infrastructure and mobility. Moreover, EPC is designed to support a variety of access technologies [19]. The rapid increase in connectivity demand has led service providers to undertake more CAPEX and OPEX investments beyond financial sense in their mobile core network infrastructure. From this point onward, it is becoming essential to have a flexible, robust, and easily manageable network; a network that could be scaled on-demand in real time and would be easily manageable. Virtualizing EPC offers all these benefits to service providers.

The basic EPC entities to support IP connectivity in LTE are the following:

| Groups | Entities | Benefits |
|---|---|---|
| Segment one | – HSS front-end (HSS FE)<br>– Mobility Management Entity (MME) | – Interactions between HSS and MME occur locally.<br>– Fewer networking transactions through Vswitches.<br>– Network transactions use the LDAP protocol, which is an efficient protocol for database information querying. |
| Segment two | – Home location register front end (HLR FE)<br>– Serving general packet radio service support node (SGSN) | – Supports combining existing SGSN with the Gn interface to the EPC system<br>– Interactions between HLR and SGSN occur locally.<br>– Fewer networking transactions through Vswitches.<br>– Network transactions use the LDAP protocol, which is an efficient protocol for database information querying. |
| Segment Three | – Packet data network gateway (PGW)<br>– Policy and charging enforcement function (PCEF)<br>– Serving gateway (SGW) | – Minimizes the number of data-plane processing nodes (flat architecture principle)<br>– Helps to overcome data-forwarding and network bottlenecks<br>– better data monitoring and charging |
| Segment Four | – User data repository (UDR).<br>– On-line charging system (OCS).<br>– Off-line charging system (OFCS).<br>– Policy and charging rules function (PCRF) | – Unified user database; less fragmentation.<br>– PCRF interacts locally with UDR to generate policies.<br>– Local interaction between OCS and PCRF<br>– Central interaction point for OSS/BSS<br>– Fewer networking transactions through Vswitches. |

*Table 2: Grouping of EPC Entities in NFV Environment.*

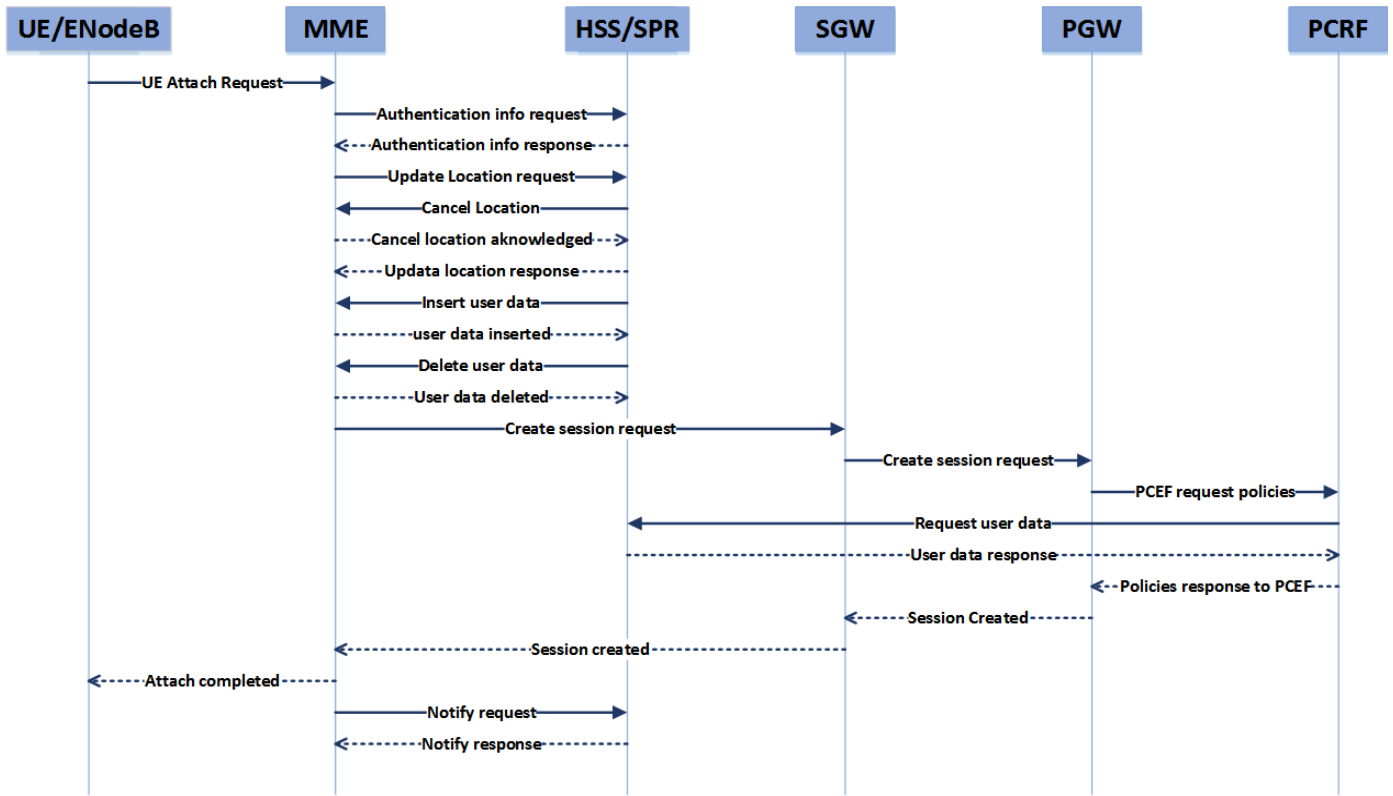

*Figure 4: Sequence Diagram for User Equipment Attachment process to LTE Network.*

1- The mobility management entity (MME) is the main control-plane entity in the LTE network.

2- The serving gateway (SGW) is responsible for routing and forwarding user data packets from and to the base station.

3- The packet data network gateway (PDN-GW) (PGW) ensures connectivity between the user data plane and external networks.

4- The Home Subscriber Server (HSS) is the central user information database.

5- The policy and charging rules function (PCRF) is responsible for passing and deciding the policies and charging in real time for each service and user.

## IX. GROUPING EPC ENTITIES IN THE NFV ENVIRONMENT

Implementing a virtualized EPC (vEPC) is the prime objective of the telecommunication equipment vendors. Since EPC encompasses multiple functionalities, instantiation of VNF in cloud has a tremendous effect on the performance and hence, VNFs are grouped together based on their interactions and workload. Generally, it is beneficial to instantiate each group in one physical server, or one local network depending on the workload.

This paper proposes an approach to vEPC entity grouping that can improve performance. The approach is based on analyzing the interconnections and functionalities of vEPC entities to achieve less control-signaling traffic and less congestion in the data plane. The proposed approach maintains the two EPC principles of flat architecture and decoupling of the control and data planes.

The grouping approach divides the entities into four segments. These are listed below and summarized in Table 2 and illustrated in Figure 5.

Since understanding the LTE framework is necessary to truly grasp the benefit of the proposed grouping, we will describe each entity's functionality as needed. Please refer to [19] for a comprehensive study of the LTE architecture.

### A. Segment One

In the proposed grouping, MME is migrated with the HSS front-end (HSS FE). The HSS front-end is an application that implements all the logical functionality of HSS but does not contain the user information database. By implementing the HSS FE with the MME, authentication and authorization processes are carried out internally, without any data transactions through the network. The HSS FE ensures that all the interactions with MME happen as if the MME was accessing the complete HSS database. The HSS FE issues a query for user information data from the user data repository (UDR), which is the central user information database and stores these data temporarily in cache memory. After querying for user information, the HSS FE acts as a complete user database and performs all authentication and authorization processes with the MME entity. Figure 4 shows the process for attaching user equipment to the LTE network. This grouping minimizes the number of network transactions that must be performed to authenticate a user because the HSS FE obtains all the required information in one query [20]. Furthermore, communication between the UDR and the HSS

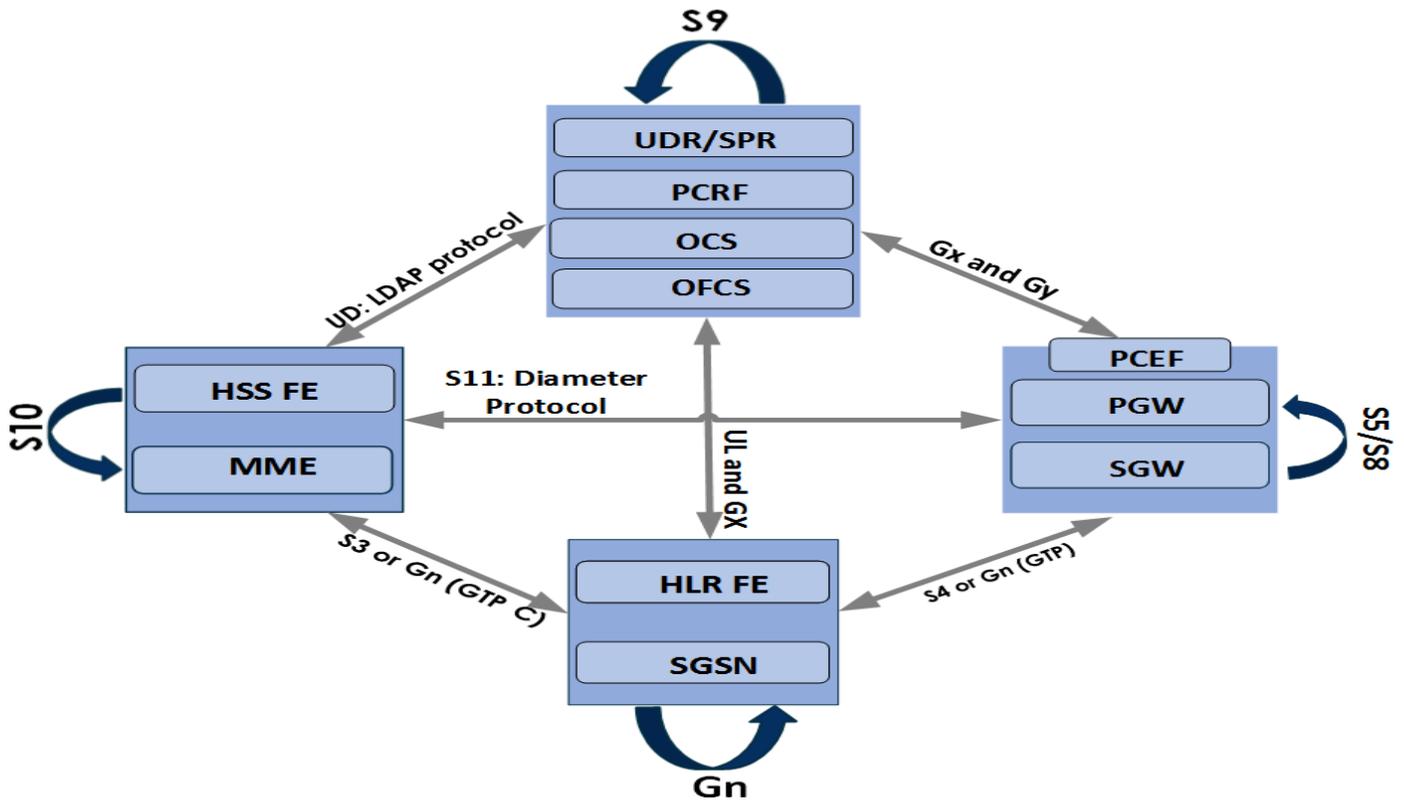

*Figure 5: vEPC Entities Grouping*

FE occurs through the lightweight directory access protocol (LDAP), not the diameter signalling protocol. LDAP is an application protocol used to exchange and manage distributed directory information services over IP networks. LDAP is a more efficient protocol than the diameter protocol for database information querying [20]. It is also faster and requires fewer resources than the diameter protocol [20-21]. It uses Transport Layer Security (TLS) or Secure Sockets Layer (SSL) to secure information exchanges, while diameter signaling uses an Internet Protocol Security (IPsec) connection for information exchanges. TLS/SSL requires fewer computational resources than IPsec and needs less initiation and resumption time. Furthermore, TLS and SSL are application-layer security protocols that provide better flexibility on a virtualized platform [22].

### B. Segment Two

In the proposed grouping, the serving general packet radio service (GPRS) support node (SGSN) is migrated with the home location register front end (HLR FE). The SGSN is a serving entity which has almost the same functionality as a combined MME and SGW. The SGSN is a network function entity existing in the GPRS core network, which permits mobile networks (2G, 3G) to transmit IP packets to external networks. It takes charge of delivering data packets to and from mobile base stations. The SGSN has user data-plane functions such as managing packet routing and transfers. Furthermore, it has control plane functions such as mobility management, logical link management, and authentication and charging functions. The SGSN is assumed in the proposed approach as almost all service providers support 2G and 3G networks besides their 4G networks. The SGSN is not combined with any EPC entities because the SGSN has a control and data plane, which contradicts the EPC architectural decoupling principle. The HLR is the database that conserves the user information in a Global System for Mobile (GSM) core network. The HLR FE is combined with the SGSN for almost the same reasons that combine the MME with the HSS FE. Moreover, this combination enables a unified data base and supports the combination of the existing SGSN with the Gn interface to the EPC system. Gn is an interface that is based on the GPRS tunneling protocol (GTP).

### C. Segment Three

In the proposed grouping, the PGW is migrated with the SGW. This merging of the two data-plane entities follows the flat architecture principle to minimize the number of data-plane processing nodes. Implementing the two entities in one VM or VNF will benefit from centralized processing in the data plane and helps to overcom the processing and network bottlenecks. In this segment, user data are not routed or transferred to the PGW after being served by the SGW. Instead, the segment has direct access to the PGW, which routes it to external networks. Centralized processing in the virtualized environment enables applications to apply the CPU affinity procedure, leading to an efficient use of CPU cache memory. In addition, this merge avoids unnecessary routing through Vswitches, which are a major bottleneck in virtual environments. Higher VNF data throughput could also be achieved using direct network interface access, in order to meet the required latencies and quality of service of the PGW and the SGW. This migration leads to a better data monitoring and charging in addition to the elimination of signaling-transaction traffic between the SGW and the PGW. All signaling transactions are carried out internally.

*D. Segment Four*

In the proposed grouping, the UDR, the PCRF, the on-line charging system (OCS), and the off-line charging system (OFCS) are migrated. Having the UDR migrated with the PCRF leads to an efficient way of generating the policy function from user information because the PCRF requests user information to generate the required policies for each established bearer. This approach prevents information exchange from overwhelming the network node, minimizes the latency of policy-function generation, and speeds policy enforcement to the PGW. As for the OCS and OFCS, the OCS is used to charge network users in a real-time manner, as in a pre-paid credit system, whereas the OFCS is used to charge users after the session is ended, as in billing services known as "pay as you go". The OCS and the OFCS interact with the PCRF and the PCEF to gather information about the session and enforce charging policies to the PGW, such as terminating the communication session when the credit limit has been exceeded. In addition, this segment groups all the entities that need to interact with the OSS/BSS. Limiting fragmentation of OSS/BSS interactions leads to more efficient control over network services.

In this grouping approach, all segments are connected almost entirely through the GPRS Tunneling Protocol (GTP), not through diameter protocol interfaces. Even though the diameter protocol is an enhanced signaling protocol in the control plane of the EPC entities, it relies on the Stream Control Transmission Protocol (SCTP) and the Transmission Control Protocol (TCP) in its transport layer. SCTP and TCP are known to downgrade network performance when small amounts of data are being exchanged [23]. These network downgrades are due to the control packets, such as acknowledgment packets, which are sent to set up the connection. When the packets are small, in general, they require more computational resources to transfer the same amount of data when they are larger. The GTP relies on the User Datagram Protocol (UDP) in its transport layer, which has satisfactory performance on small-packet data-exchange connections [23]. Because control-signaling packets are small, using an approach that maintains interfaces on GTP leads to better computing performance and use of network resources. Although the proposed grouping introduces benefits in terms of minimizing control signaling traffic to avoid congestion on the networking infrastructure, it requires much computational power because most transactions are carried out internally.

| Traffic Profile | |
|---|---|
| Registered Subscribers | 167,650 |
| Subscribers Attached to The Network | 150,878 |
| Busy Hour Session Attempts | 64,940,898 |
| Simultaneous Evolved Packet System Bearers | 18,853 |
| **Planning Parameters** | |
| Mean Session Time | 180 sec |
| Handover Ratio | 0.4 |
| Dense Area Attached Subscriber Ratio | 0.9 |
| Average EPSB session duration | 900 sec |
| Busy Hour Traffic Ratio | 0.15 |
| Retransmission factor | 0.25 |
| Pre-paid Accounts | 80% |

*Table 3: Traffic profile and Planning parameters [24].*

## X. QUANTITATIVE ANALYSIS

The main intent of the proposed grouping is reducing control-signaling traffic in the EPC, which is expected to increase exponentially by 2015 [24]. To illustrate the proposed grouping benefits, the signaling traffic generated in [25] were applied, and the reduction in the required bandwidth is estimated. In [25], the signaling traffic was generated for 15 eNBs connected to the EPC entities where traffic profile and planning parameters are shown in Table 3. However, the total signaling transaction traffic between the MME and the HSS in [25] was 1,039,430 transactions, and the average number of transactions per second between the MME and the HSS was 6.2 transactions per subscriber. Using the proposed grouping of MME and HSS FE, the number of transactions decreased from 6.2 to 1 transaction(s) per subscriber. The reason behind this reduction in the number of transactions is the combination of all the user information in one query from the UDR [25]. Consequently, total transaction traffic was reduced to 173,239 transactions per second. In [25], signaling transaction traffic between the SGW and the PGW was 56,559 transactions. Using the proposed grouping of the SGW and the PGW, signaling transactions over the network were eliminated. Moreover, total signaling-transaction traffic for PCRF was 37,706, with an average of two transactions per bearer in [25]. As a result, the number of PCRF and UDR signaling transactions was 18,853. Because 80 percent of users had pre-paid accounts, 30,164 transactions between the PCRF and the OCS were generated. Using the proposed grouping, these transactions were eliminated because the PCRF, OCS, and UDR are implemented in the same segment. Signaling traffic takes place between these entities and the PCEF, which is implemented with the PGW in a different segment. These results are illustrated in Table 4.

| Transactions between Core Elements | Signaling (transactions per sec) [24] | After Grouping |
|---|---|---|
| MME, eNBs , and S-GW | 175,332 | 175,332 |
| S-GW and P-GW | 56,559 | 0 (Internal transactions) |
| MME and HSS | 1,039,430 | 173,239 to UDR |
| PCRF and P-GW | 37,706 | 37,706 |
| PCRF and UDR | 18853 | 0 (Internal transactions) |
| PCRF and OCS | 30164 | 0 (Internal transactions) |
| Total Traffic | 1358044 | 386277 |

*Table 4: Signalling traffic before and after grouping.*

## XI. CONCLUSIONS

The NFV aims to revolutionize the telecommunication industry by decoupling network functions from the underlying proprietary hardware. It provides all the benefits of IT virtualization platforms. Academic researchers and network engineers are exploiting virtual environments to simplify and enhance NFV in order to find its way smoothly into the telecommunications industry. Besides all the advantages brought by NFV to the telecommunications industry, it faces technical challenges that might hinder its progress. Therefore, IT organizations, network enterprises, telecommunication equipment vendors, and academic researchers should be aware of these challenges and explore new approaches to overcome them.

This work also proposed a grouping criterion for virtualized network functions to minimize the transaction occurred on the physical network. The analysis shows that the proposed grouping can reduce the network control traffic by 70 percent.

## ACKNOWLEDGMENT

We are indebted to Mr. Peter Willis from British Telecom (BT) for his support and valuable discussions.

## BIOGRAPHIES

**Hassan Hawilo** (hhawilo@uwo.ca) received his B.E. degree in communication and electronics engineering in 2012 from Beirut Arab University, Lebanon. He is currently working towards the M.E.Sc. degree in cloud computing and virtualization technologies at Western Ontario University, Canada. His research interests include cloud computing, virtualization, software defined network, Network Function Virtualization, and highly available software.

**Abdallah Shami** received his B.E. degree in Electrical and Computer Engineering from the Lebanese University, Beirut, Lebanon in 1997, and the Ph.D. Degree in Electrical Engineering from the Graduate School and University Center, City University of New York, New York, NY in September 2002. Since July 2004, he has been with Western University, Canada where he is currently a Professor in the Department of Electrical and Computer Engineering. His current research interests are in the area of network-based cloud computing and wireless/data networking. Dr. Shami has chaired key symposia for IEEE GLOBECOM, IEEE ICC, IEEE ICNC, and ICCIT. Dr. Shami is a Senior Member of IEEE and elected Chair of the IEEE Communications Society Technical Committee on Communications Software.

**Maysam Mirahmadi** received the B.Sc. and M.Sc. degrees in Electrical Engineering from Amirkabir University of Technology (Tehran Polytechnic), Tehran, Iran, in 2004 and 2007, respectively. From 2007 to 2009, he was a Senior Designer at Kavoshcom R&D Group, Tehran, Iran. In 2013, he received the Ph.D. degree from the University of Western Ontario, London, Ontario, Canada. Since 2013, he has been with IBM Canada Research and Development Center as research scientist. His current research interests include wireless communication, heterogeneous computing, and cloud computing.

**Dr Rasool Asal** is a Chief Researcher at Etisalat BT Innovation Centre (EBTIC) leading EBTIC research and innovation activities in the area of Cloud Computing. His current interest focuses primarily on the Cloud Technologies, Cloud Security Architectures and the design of wide-area distributed cloud compliance enterprise systems that scale to millions of users. For the past fifteen years, he has been working with British Telecommunications Group at Adastral Park (Ipswich, U.K), designing and developing a considerable volume of high-performance enterprise applications, mostly in the area of telecommunications.